\documentstyle[12pt]{article}

\newcommand{\eq}{\begin{equation}}
\newcommand{\eqn}{\begin{displaymath}}
\newcommand{\en}{\end{equation}}
\newcommand{\enn}{\end{displaymath}}
\def\pnot{\mbox{${\not{\hbox{\kern-3.0pt$p$}}}$}}
\def\qnot{\mbox{${\not{\hbox{\kern-2.0pt$q$}}}$}}
\def\enot{\mbox{${\not{\hbox{\kern-2.0pt$e$}}}$}}
\def\knot{\mbox{${\not{\hbox{\kern-2.0pt$k$}}}$}}
\def\rnot{\mbox{${\not{\hbox{\kern-2.0pt$r$}}}$}}
\def\r'not{\mbox{${\not{\hbox{\kern-2.0pt$r'$}}}$}}
\def\la{\mathrel{\mathpalette\fun <}}

\def\fun#1#2{\lower3.6pt\vbox{\baselineskip0pt\lineskip.9pt\ialign
{$\mathsurround=0pt#1\hfil##\hfil$\crcr#2\crcr\sim\crcr}}}

\newcommand{\lsim}{\ \raise -2.truept\hbox{\rlap{\hbox{$\sim$}}
\raise5.truept \hbox{$<$}\ }}
\newcommand{\gsim}{\ \raise -2.truept\hbox{\rlap{\hbox{$\sim$}}
\raise5.truept\hbox{$>$}\ }}

\begin{document}

\begin{titlepage}
\hskip 12cm \vbox{\hbox{BUDKERINP/95-49}\hbox{CS-TH 12/95}\hbox{June 1995}}
\vskip 0.6cm
\centerline{\bf REGGEIZATION OF QUARK-QUARK SCATTERING}
\centerline{\bf AMPLITUDE IN QCD$^{~\ast}$}
\vskip 1.0cm
\centerline{  V.Fadin$^{\dagger}$}
\vskip .5cm
\centerline{\sl Budker Institute for Nuclear Physics}
\centerline{\sl and Novosibirsk State University, 630090 Novosibirsk,
Russia}
\vskip .5cm
\centerline{  R.Fiore$^{\ddagger}$, A.Quartarolo$^{\ddagger}$}
\vskip .5cm
\centerline{\sl  Dipartimento di Fisica, Universit\`a della Calabria}
\centerline{\sl Istituto Nazionale di Fisica Nucleare, Gruppo collegato di
Cosenza}
\centerline{\sl Arcavacata di Rende, I-87030 Cosenza, Italy}
\vskip 1cm
\begin{abstract}
$s$-channel discontinuity of quark-quark scattering amplitude with
gluon quantum numbers in the $t$ channel and negative signature is
calculated in the Regge kinematical region in the two-loop approximation.
Using this discontinuity and assuming that the Regge asymptotic behaviour is
given by the Reggeized gluon contribution, we calculate the gluon
trajectory in the two-loop approximation. Remarkable cancellations lead to
the independence of the trajectory on properties of the scattered quarks,
confirming the gluon Reggeization.
\end{abstract}
\vskip .5cm
\hrule
\vskip.3cm
\noindent

\noindent
$^{\ast}${\it Work supported in part by ISF under grant RAK-000, in part by
the Ministero italiano dell'Universit\`a e della Ricerca Scientifica e
Tecnologica and in part by the EEC Programme ``Human Capital and Mobility",
Network ``Physics at High Energy Colliders", contract CHRX-CT93-0537 (DG 12
COMA). INTAS grant 93-1867}
\vfill

$ \begin{array}{ll}
^{\dagger}\mbox{{\it email address:}} &
 \mbox{FADIN~@INP.NSK.SU}\\
\end{array}
$

$ \begin{array}{ll}
^{\ddagger}\mbox{{\it email address:}} &
  \mbox{FIORE, QUARTAROLO~@FIS.UNICAL.IT}
\end{array}
$
\vfill
\end{titlepage}
\eject
\baselineskip=24pt
{\bf 1. INTRODUCTION}

Perturbative QCD is widely used nowadays for the description of semihard
{}~\cite{GLR} as well as hard processes~\cite{AL}. However, whereas for the
later ones the theory is well developed and understood, for the former ones we
have a lot of unsolved theoretical problems. The applicability of
perturbation theory improved by the renormalization group for a hard process
having a large typical virtuality $Q^2$ is justified by the smallness of the
strong coupling constant $\alpha_s(Q^2)$. Contrary, for semihard processes
another essential parameter appears: the ratio of the typical virtuality
$Q^2$ to the square of the
c.m.s energy $s$ of colliding particles. At sufficiently high energy the
value $x=Q^2/s$ becomes so small that it is necessary to sum up
terms of the type $\alpha_s^n(ln(1/x))^m$, with $m\leq n$
(for the
scattering channel which is considered here). Up to now this problem is
solved~\cite{FKL} in the leading logarithmic approximation (LLA) only, which
means summation of the terms with $m=n$.
\vskip.3cm
The results of LLA have two serious disadvantages. Firstly, the Froissart
bound ${\sigma_{tot}}<c~{\ln^2{s}}$ is violated in LLA. In fact,
calculated in LLA,
the total cross section ${\sigma}^{LLA}_{tot}$ grows at large c.m.s.
energies as a power of $s$:
\eq
{\sigma}^{LLA}_{tot}\sim {s^{\omega_{0}}\over {\sqrt{\ln{s}}}}~,
\label{z1}
\en
where, for the gauge group SU({N}) ($N=3$ for QCD), with gauge coupling
constant $g$
($\alpha_s={g^2\over 4\pi}$),
\eq
{\omega_{0}} = {g^2\over \pi^2}N{\ln{2}}~.
\label{z2}
\en
In terms of structure functions this means their strong power increase in
the small $x$ region. The Froissart bound is violated in LLA because the
$s$-channel unitarity constraints for scattering amplitudes are not
completely fulfilled in this approximation. The problem of
{\it unitarization} of LLA results is extremely important from a
theoretical point of view. It is concerned in a lot of papers (see, for
example, Ref.~\cite{MLBCN}).
\vskip.3cm
Another disadvantage seems to be even more important from a practical point
of view, since the results of LLA are applied to the small $x$
phenomenology (see, for instance, Ref.~\cite{MNK}). There is an uncertainty
of the argument of the running coupling constant which appears because the
scale dependence of $\alpha_s$ is beyond of the accuracy of LLA.
This uncertainty diminishes the predictive power of LLA, permitting to change
strongly numerical results by changing a scale.
\vskip.3cm
Therefore, the problem of the calculation of radiative corrections to
LLA becomes very important now, as it gives us the possibility to fix the
scale dependence of the coupling constant, to reduce the uncertainty of the
predictions of LLA and to determine a region of its applicability.
\vskip.3cm
A solution of this problem can be strongly simplified \cite{LF} by using the
Reggeization property of the non-Abelian SU({N}) gauge theories. It was
proved \cite{FKL,BLF} in LLA that gauge bosons are Reggeized in these
theories with trajectory
\eq
j(t) = 1+\omega(t)~,
\label{z3}
\en
where in the leading approximation
\eq
\omega(t) = \omega^{(1)}(t) = {g^2t\over {(2{\pi})}^{(D-1)}}\frac{N}{2}
\int{d^{D-2}k_\perp\over k_\perp^2{(q-k)}_{\perp}^2}~.
\label{z4}
\en
Here $q$ is the momentum transfer, $t=q^2\approx q_\perp^2$, and
$D=4+\varepsilon$ is the space-time dimension. A non zero $\varepsilon$ is
introduced to regularize Feynman integrals. The integration in Eq.(\ref{z4})
is performed over the ($D-2$)-dimensional momenta orthogonal to the initial
particle momentum plane.
\vskip.3cm
The results of LLA are summarized \cite{FKL} in the Bethe-Salpeter type
equation for the $t$-channel partial amplitudes. The problem of the
calculation of corrections to LLA, therefore, can be set up as the
problem of the calculation of corrections to the kernel of this equation
\cite{LF}. The kernel is expressed in terms of the gluon trajectory and the
Reggeon-Reggeon-gluon (RRG) vertex. The corrections to the vertex are known
\cite{FL,FFQ2}, therefore the calculation of the contribution
$\omega^{(2)}(t)$ to the trajectory in the next (two-loop) approximation
has become most urgent.
\vskip.3cm
In this paper we present results and details of the calculation of the
two-loop correction to the trajectory for the real case of QCD with massive
quark flavours. The result for $\omega^{(2)}(t)$ in the massless quark case
was obtained earlier \cite{VF}. The paper is organized as follows. In Sec.
II we discuss the method of calculation. In Sec. III we calculate the
contribution of the two-particle intermediate state to the $s$-channel
discontinuity of the quark-quark scattering amplitude. An analogous
calculation is performed in Sec. IV for the contribution of the three particle
intermediate state. The final expression for the correction $\omega^{(2)}(t)$
is obtained and discussed in Sec. V.

\vskip 0.3cm

{\bf 2. METHOD OF CALCULATION}

The method is based on using $s$-channel unitarity. The two-loop
contribution to the gluon trajectory can be obtained from the $s$-channel
discontinuity of an elastic scattering amplitude with gluon quantum numbers
and negative signature in the $t$ channel calculated in the two-loop
approximation with the accuracy up to a constant. Indeed, let us consider
such an amplitude for a process of the type $A+B\rightarrow A'+B'$ at large
$s$ and fixed $t$. Assuming the gluon Reggeization, the amplitude takes the
factorized form
\eq
\left({{\cal A}_8}^{(-)}\right)^{A'B'}_{AB} = \Gamma^i_{A'A}{s\over t}
\left[\left({s\over -t}\right)^{\omega(t)}+\left({-s\over -t}\right)^
{\omega(t)}\right]\Gamma^i_{B'B}~,
\label{z5}
\en
where $\Gamma^i_{A'A}$ are the particle-particle-Reggeon (PPR) vertices.
They can be written as
\eq
\Gamma^{i}_{A'A} = g\langle A'| T^i| A \rangle (\Gamma^{(0)}_{A'A}+
\Gamma^{(1)}_{A'A})~,
\label{z6}
\en
where $\langle A'| T^i| A \rangle$ stands for a matrix element of the
colour group generator in the corresponding representation (i.e. fundamental
for quarks and adjoint for gluons), $\Gamma^{(0)}_{A'A}$ and
$\Gamma^{(1)}_{A'A}$ are respectively the Born and the one-loop contributions
to the vertices. Using Eq.(\ref{z6}) and the decomposition
\eq
\omega(t) = \omega^{(1)}(t)+\omega^{(2)}(t)
\label{z7}
\en
one can present the two-loop contribution to the $s$-channel discontinuity
$\left[\left({{\cal A}_8}^{(-)}\right)^{A'B'}_{AB}\right]_S$ in the form
\eqn
\left[\left({{\cal A}_8}^{(-)}\right)^{A'B'}_{AB}(\mbox{two-loop})\right]_S =
(-\frac{2\pi is}{t})g^2\langle A'| T^i| A \rangle\langle B'| T^i| B \rangle
\enn
\eq
\times \left[\Gamma^{(0)}_{A'A}(\omega^{(1)}(t))^2\ln{(\frac{s}{-t})}
\Gamma^{(0)}_{B'B}+(\Gamma^{(1)}_{A'A}\Gamma^{(0)}_{B'B}+
\Gamma^{(0)}_{A'A}\Gamma^{(1)}_{B'B})\omega^{(1)}(t)+
\Gamma^{(0)}_{A'A}\omega^{(2)}(t)\Gamma^{(0)}_{B'B}\right]~.
\label{z8}
\en
Since the one-loop corrections $\Gamma^{(1)}_{A'A}$ to the PPR vertices
became available \cite{FL,FF,FFQ1}, the only unknown quantity in the RHS of
Eq.(\ref{z8}) is the two-loop contribution $\omega^{(2)}(t)$ to the gluon
trajectory. Consequently, one may obtain it from the expression of the
discontinuity $\left[\left({{\cal A}_8}^{(-)}\right)^{A'B'}_{AB}
(\mbox{two-loop})
\right]_S$ calculated in the two-loop approximation with accuracy up to a
constant. The calculation of this discontinuity is the main content of
this paper.
\vskip.3cm
By definition the trajectory should not depend on a particular type
of scattered particles, therefore we have a freedom in choosing these
particles. In this paper we consider the process of the quark-quark
scattering for calculating the correction to the trajectory.
\vskip.3cm
In LLA helicities of each of the scattered particles are conserved, so
that in the helicity basis we have
\eq
\Gamma^{(0)}_{A'A} = \delta_{{\lambda_{A'}}{\lambda_A}}~.
\label{z9}
\en
On the contrary, in higher orders PPR vertex $\Gamma^{(i)}_{A'A}$ may
contain another spin structure. Because of the parity conservation, the
one-loop correction  $\Gamma^{(1)}_{A'A}$ can be written as \cite{FFQ1}
\eq
\Gamma^{(1)}_{A'A} = \delta_{{\lambda_A},{\lambda_{A'}}}\Gamma^{\left(+
\right)}_{AA}(t)+\delta_{{\lambda_A},{-\lambda_{A'}}}\Gamma^{\left(-\right)}_
{AA}(t)
\label{z10}
\en
if the relative phases of the states with opposite helicity are appropriately
chosen. Both quantities $\Gamma^{\left({\pm}\right)}_{AA}(t)$ are calculated
in Ref.~\cite{FFQ1}. For our purpose the knowledge of $\Gamma^{\left(+
\right)}_{AA}(t)$ is sufficient. In fact, the correction $\omega^{(2)}(t)$ in
Eq.(\ref{z8}) is multiplied by the helicity conserving vertices, then for
its determination only the helicity conserving parts of the amplitudes
entering Eq.(\ref{z8}) are necessary.
\vskip.3cm
Let us write the helicity conserving part of the discontinuity of the
amplitude in the LHS of Eq.(\ref{z8}) in the following form:
\eq
\left[\left({{\cal A}_8}^{(-)}\right)^{A'B'}_{AB}(\mbox{two-loop})\right]^
{(+)}_S =
g^2\langle A'|T^i| A \rangle\langle B'|T^i| B \rangle
\left(-\frac{2{\pi}is}{t}
\right)\Delta_s~,
\label{z11}
\en
where the superscript $(+)$ in the LHS means {\it helicity conserving part}.
Then Eq.(\ref{z8}), together with Eqs.(\ref{z9}) and (\ref{z10}), gives us
\eq
\omega^{(2)}(t) = \Delta_s-\left(\omega^{(1)}(t)\right)^2\ln\left(\frac
{s}{-t}\right)-\left(\Gamma^{\left(+\right)}_{AA}(t)+\Gamma^{\left(+\right)}_
{BB}(t)\right)\omega^{(1)}(t)~.
\label{z12}
\en
The only unknown term in the RHS of Eq.(\ref{z12}) is the discontinuity
$\Delta_s$. We calculate it below by using the $s$-channel unitarity
condition. In the two-loop approximation only two- and three-particle
intermediate states do contribute. This allows to divide
the discontinuity under consideration into two parts respectively:
\eq
\Delta_s = \Delta^{(2)}_s+\Delta^{(3)}_s~.
\label{z13}
\en

\vskip 0.3cm

{\bf 3. TWO-PARTICLE CONTRIBUTION TO THE DISCONTINUITY}

Let us consider the two-particle contribution in the unitarity relation:
\eq
\left[{\cal A}^{A'B'}_{AB}\right]_{S(2)} =
i\int{d\Phi_2\left(p_A+p_B;p_{A_1},p_{B_1}\right)\sum_{{A_1},{B_1}}{\cal A}
^{A_1B_1}_{AB}{\cal A}^{\ast A_1B_1}_{A'B'}}~.
\label{z14}
\en
Since we are considering the quark-quark scattering amplitude, the
intermediate particles can be only quarks and the summation is
performed over their spin and colour states; $d\Phi_2$ is the two-body
phase space element. Its general expression reads
\eq
d\Phi_n\left(P;p_1,....,p_n\right) = (2\pi)^D\delta^{(D)}\left(P-
\sum_{i=1}^np_i\right)\prod_{i=1}^n\frac{d^{(D-1)}p_i}{2E_i(2\pi)^{D-1}}~.
\label{z15}
\en
\vskip.3cm
In order to obtain the two-loop contribution to the discontinuity we need to
take one of the two amplitudes in the RHS of Eq.(\ref{z14}) in the Born
approximation and the other one in the one-loop approximation (see Fig. 1).
Both amplitudes should be calculated in the region of asymptotically large
c.m.s. energies $\sqrt {s}$ and fixed momentum transfers. In this region
the amplitudes acquire a very simple structure in the Born approximation:
\eq
{\cal A}^{A_1B_1}_{AB}(\mbox {Born}) = \langle A_1| T^i| A \rangle\langle B_1|
T^i| B \rangle\delta_{{\lambda_A},{\lambda_{A_1}}}\delta_{{\lambda_B},
{\lambda_{B_1}}}\frac{2g^2s}{t_1}~,
\label{z16}
\en
where $t_1=(p_A-p_{A_1})^2$. The amplitude ${\cal A}^{A_1B_1}_{A'B'}$ may be
obtained from Eq.(\ref{z16}) by evident substitutions.
\vskip.3cm
The one-loop contributions to the amplitudes ${\cal A}^{A_1B_1}_{AB}$ and
${\cal A}^{A_1B_1}_{A'B'}$ have a much more complicated form. Fortunately, it
is sufficient to take into account only those parts of these contributions
which conserve the helicity of each particle and have negative signature
in $t_1$ and $t_1'$ channels correspondingly $\left(t_1'=(p_{A'}-p_
{A_1})^2\right)$. Indeed, let us consider the spin structure. We are
interested in the helicity conserving part of the discontinuity, so that
only such a part should be kept in the product of the Born and one-loop
contributions in Eq.(\ref{z14}). Since in the Born approximation the
helicities are conserved, as Eq.(\ref{z16}) shows, the only helicity
conserving part must be taken in the one-loop contribution.
Now let us turn to the colour structure and signature.
In the Born approximation
(\ref{z16}) the amplitudes evidently have gluon quantum numbers in $t_1$
and $t_2$ channels and negative signature (in the region of asymptotically
large $s$ it simply means that they are odd functions of $s$).  The
one-loop contribution to the amplitudes ${\cal A}^{A_1B_1}_{AB}$ and
${\cal A}^{A_1B_1}_{A'B'}$, besides such a part, has also a positive signature
part, with colour singlet as well as octet state in the $t_1$ and $t_1'$
channels. This part, however, is purely imaginary \cite{FFQ1}, whereas the
Born amplitude is real; therefore, their product cancel in the unitarity
equation (\ref{z14}).
\vskip.3cm
As for the result, to calculate the discontinuity $\Delta^{(2)}_S$ we may use,
for the amplitudes ${\cal A}^{A_1B_1}_{AB}$ and ${\cal A}^{A_1B_1}_{A'B'}$
in Eq.(\ref{z14}), the representation (\ref{z5}) with the trajectory
$\omega(t)$ and vertices $\Gamma^c_{A'A}$ given by Eq.(\ref{z4}) and Eqs.
(\ref{z6}), (\ref{z9}) and (\ref{z10}) respectively. In the one-loop
approximation we get
\eqn
{\cal A}^{A_1B_1}_{AB}(\mbox{one-loop}) = \langle A_1| T^i| A \rangle
\langle B_1| T^i| B \rangle\delta_{{\lambda_A},{\lambda_{A_1}}}
\delta_{{\lambda_B},{\lambda_{B_1}}}
\enn
\eq
\times\frac{g^2s}{t_1}\left\{\omega^{(1)}(t_1)\left[\ln{\left(\frac{s}{-t_1}
\right)}+\ln{\left(\frac{-s}{-t_1}\right)}\right]+
2\left[\Gamma^{(+)}_{AA}(t_1)+\Gamma^{(+)}_{BB}(t_1)\right]
\right\}+\dots~,
\label{z17}
\en
where dots denote terms which do not contribute to the discontinuity
$\Delta^{(2)}_s$.
The amplitude ${\cal A}^{A_1B_1}_{A'B'}(\mbox{one-loop})$ is obtained from
Eq.(\ref{z17}) by evident substitutions. Using the Born (\ref{z16}) and
one-loop (\ref{z17}) amplitudes and summing
up in Eq.(\ref{z14}) over spin and colour states of the particles $A_1$ and
$B_1$ leads to the spin and colour factor
\eqn
\langle A'| T^iT^j| A \rangle\langle B'| T^iT^j| B \rangle\delta_{{\lambda_A},
{\lambda_{A'}}}\delta_{{\lambda_B},{\lambda_{B'}}}.
\enn
Let us remind that we need to calculate not the discontinuity (\ref{z14})
itself, but the corresponding discontinuity for the amplitude with the
octet colour state in the $t$ channel and negative signature. The calculation
of the discontinuity for this amplitude requires antisymmetrization with
respect to the change $A \rightarrow \bar A'$, $A' \rightarrow \bar A$ or,
equivalently, $B \rightarrow \bar B'$, $B' \rightarrow \bar B$, that in our
case reduces to the change of the order of the group generators into the
factor $\langle A'| T^iT^j| A \rangle$. Using the relations
\eq
\left[T^i,T^j\right] = if^{ij}_{~~~k}T^k~,~~~~~~~~if_{ij}^{~~~k}T^iT^j =
-\frac{N}{2}T^k~,
\label{z18}
\en
where the coefficients $f^{ijk}$ are the group structure constants, the colour
factor becomes
\eqn
-\frac{N}{4}\langle A'| T^i| A \rangle\langle B'| T^i| B \rangle~.
\enn
Notice that the negative signature automatically leads here to the colour
octet state in the $t$ channel. Using the two-body phase space element
\eq
d\Phi_2\left(p_A+p_B;p_{A_1},p_{B_1}\right) = \frac{1}{2s}\frac{d^{(D-2)}
q_{1\perp}}{(2\pi)^{D-2}}~,
\label{z19}
\en
with $q_1=p_{A_1}-p_{A}$, from Eqs.(\ref{z14}), (\ref{z16}) and (\ref{z17})
we get
\eqn
\left[\left({{\cal A}_8}^{(-)}\right)^{A'B'}_{AB}(\mbox{two-loop})\right]^
{(+)}_{S(2)} =
g^2\langle A'| T^i| A \rangle\langle B'| T^i| B \rangle
\left(-\frac{2{\pi}is}{t}\right)\frac{g^2Nt}{(2\pi)^{D-1}}
\enn
\eq
\times\int\frac{d^{(D-2)}q_{1\perp}}
{(q_1-q)^2_{\perp}q^2_{1\perp}}
\left[\omega^{(1)}(q^2_{1\perp})\ln{\left(\frac{s}{-q^2_{1\perp}}\right)}+
\Gamma^{(+)}_{AA}(q^2_{1\perp})+\Gamma^{(+)}_{BB}(q^2_{1\perp})\right]~.
\label{z20}
\en
Comparing Eq.(\ref{z20}) with Eqs.(\ref{z11}) and (\ref{z13}),
we obtain
\eqn
\Delta^{(2)}_s =
\enn
\eq
\frac{g^2Nt}{(2\pi)^{D-1}}\int\frac{d^{(D-2)}q_
{1\perp}}{q^2_{1\perp}(q_1-q)^2_{\perp}}
\left[\omega^{(1)}(q^2_{1\perp})\ln{\left(\frac{s}{-q^2_{1\perp}}\right)}+
\Gamma^{(+)}_{AA}(q^2_{1\perp})+\Gamma^{(+)}_{BB}(q^2_{1\perp})\right]~.
\label{z21}
\en
The helicity conserving part of the one-loop correction to the
quark-quark-Reggeon vertex $\Gamma^{(+)}_{QQ}(t)$ is calculated in Ref.~
\cite{FFQ1}. For our purpose it is convenient to express it as a sum of
three parts having different flavour-colour dependence:
\eq
\Gamma^{(+)}_{QQ}(t) = a_f(t)+a_Q(t,m^2_Q)+a_g(t,m^2_Q)~.
\label{z22}
\en
Here the first term
\eq
a_f(t) = -2{g^2\over{(4\pi)^{D\over2}}}\Gamma\left(2-{D\over 2}\right)
\sum_{f}\int^1_0dx{x(1-x)\over{\left[m_f^2-x(1-x)t\right]^{2-{D\over 2}}}}
\label{z23}
\en
is the contribution of the quark loop, the summation being over the quark
flavours. The second term
\eqn
a_Q(t,m^2_Q) = {g^2\over{(4\pi)^{D\over2}}}\frac{1}{2N}\Gamma\left(2-
{D\over 2}\right)\left\{ \vbox to 24pt{}\int_0^1
{dx\over{\left[m^2_Q-x(1-x)t\right]}^{3-{D\over 2}}}\right.
\enn
\eq
\left.\times\left[t\left({1\over D-3}+\frac{D-4}{4}\right)-
\frac{2m^2_Q}{(D-3)}\right]+\frac{2}{D-3}(m^2_Q)^{\frac{D}{2}-2}
\vbox to 24pt{}\right\}
\label{z24}
\en
is connected with the vertex and the quark self-energy
diagrams. Eq.(\ref{z24}) is obtained  from Eqs.(39) and (68)
of Ref.~\cite{FFQ1} with the help of the identity
\eq
(D-3)\int_0^1dx\frac{tx(1-x)}{\left[m^2-x(1-x)t\right]^{3-{D\over 2}}} =
\frac{1}{4}\int_0^1dx\frac{t(D-4)+4m^2}{\left[m^2-x(1-x)t\right]^
{3-{D\over 2}}}-\left(m^2\right)^{\frac{D}{2}-2}~,
\label{z25}
\en
which in turn follows from the evident identity
\eq
\int_0^1dx\frac{d}{dx}\left(\frac{x}{\left[m^2-x(1-x)t\right]^{2-{D\over 2}}}
\right) = \left(m^2\right)^{\frac{D}{2}-2}~.
\label{z26}
\en
Finally, the last term, proportional to N, comes from the two gluon exchange
and quark self-energy diagrams; it can be written as
\eq
a_g(t,m^2_Q) = a_g(t,0)+\delta_g(t,m^2_Q)~.
\label{z27}
\en
Here $a_g(t,0)$ is given by Eq.(63) of Ref.~\cite{FFQ1},
\eqn
a_g(t,0) = {g^2N\over \left(4\pi\right)^{D\over 2}}{\Gamma\left(2-{D\over 2}
\right)\Gamma^2\left({D\over 2}-1\right)\over \left(-t\right)^{2-{D\over 2}}
\Gamma\left(D-2\right)}\left\{\vbox to 16.66pt{}\left(D-3\right)
\left[\psi \left(3-{D\over 2}\right)\right.\right.
\enn
\eq
\left.\left. -2\psi \left({D\over 2}-2\right)+\psi \left(1\right)\right]+
{1\over 4\left(D-1\right)}-{2\over D-4}-{7\over 4}\right\}~,
\label{z28}
\en
where $\psi(x)$ is the logarithmic derivative of the gamma
function:
\eqn
\psi(x) = \frac{\Gamma'(x)}{\Gamma(x)}~.
\enn
In turn, the second term $\delta_g(t,m^2_Q)$ comes from Eq.(64) and from the
part proportional to N in Eq.(67) of Ref.~\cite{FFQ1}. It is
convenient to modify its form using the identity
\eqn
\int^1_0\int^1_0{dx_1dx_2\theta(1-x_1-x_2)\left(\frac{\partial}{\partial x_2}
-2\frac{\partial}{\partial x_1}\right)\frac{x_1}{\left[m^2x_1^2-tx_2
\left(1-x_1-x_2\right)\right]^{2-\frac{D}{2}}}} =
\enn
\eq
-\frac{2}{D-2}(m^2)^{\frac{D}{2}-2}~,
\label{z29}
\en
which provides us the relation
\eqn
2\int^1_0\int^1_0{dx_1dx_2\frac{tx_2(1-x_1-x_2)\theta(1-x_1-x_2)}{\left[m^2
x_1^2-tx_2\left(1-x_1-x_2\right)\right]^{3-\frac{D}{2}}}}+\frac{2}{D-2}
(m^2)^{\frac{D}{2}-2} =
\enn
\eq
\int^1_0\int^1_0{dx_1dx_2\frac{\theta(1-x_1-x_2)\left[\left(\frac{D}{2}
-2\right)tx_1(1-x_1)+2(D-3)m^2x^2_1\right]}{\left[m^2x_1^2-tx_2
\left(1-x_1-x_2\right)\right]^{3-\frac{D}{2}}}}~.
\label{z30}
\en
Using this relation one arrives at
\eqn
\delta_g(t,m^2_Q) = \frac{g^2N}{\left(4\pi\right)^{D\over 2}}\Gamma\left(3-
{D\over 2}\right)\left\{ \vbox to 24pt{}\int^1_0\int^1_0{dx_1dx_2
\theta(1-x_1-x_2)}\right.
\enn
\eqn
\left.\times\left[\frac{t(1-x_1)}{x_1}\left(1-x_1+\frac{D-2}{4}x^2_1
\right)\left(\vbox to 24pt{}
\frac{1}{\left[m^2_Qx_1^2-tx_2\left(1-x_1-x_2\right)
\right]^{3-\frac{D}{2}}}\right.\right.\right.
\enn
\eqn
\left.\left.\left.-\frac{1}{\left[-tx_2\left(1-x_1-x_2\right)\right]^
{3-\frac{D}{2}}}\vbox to 24pt{}\right)-
\frac{2m^2_Qx_1}{\left[m^2_Qx_1^2-tx_2\left(1-x_1-
x_2\right)\right]^{3-\frac{D}{2}}}\right]\right.
\enn
\eq
\left. +\frac{2(m^2_Q)^{\frac{D}{2}-2}}{(D-3)(D-4)}\vbox to 24pt{}\right\}
\label{z31}
\en

\vskip 0.3cm

{\bf 4. THREE-PARTICLE CONTRIBUTION TO THE DISCONTINUITY}

Let us now turn to the calculation of the contribution coming from the
three-particle intermediate
state in the $s$-channel unitarity condition. Since we are considering the
discontinuity of the quark-quark scattering amplitude, in the intermediate
state we face the same quarks with an additional gluon. The unitarity
condition in this case reads
\eq
\left[{\cal A}^{A'B'}_{AB}\right]_{S(3)} =
i\int{d\Phi_3\left(p_A+p_B;p_{A_1},p_G,p_{B_1}\right)\sum_{{A_1},G,{B_1}}
{\cal A}^{A_1GB_1}_{AB}{\cal A}^{\ast A_1GB_1}_{A'B'}}~.
\label{z32}
\en
where $\sl G$ is the produced gluon. For calculating the
discontinuity it is convenient to use Sudakov variables. Let us introduce
light-like momenta $p_{-}$ and $p_{+}$ close to the initial momenta $p_A$
and $p_B$ respectively:
\eqn
p_A = p_{-}+\frac{m^2_A}{s}p_{+}~,~~~~~~p_B = p_{+}+\frac{m^2_B}{s}p_{-}~,
\enn
\eq
p^2_{-} = p^2_{+} = 0~,~~~~~~2p_{+}p_{-} = s~.
\label{z33}
\en
Thus any vector $p_i$ can be decomposed as
\eq
p_i = \beta_ip_{-}+\alpha_ip_{+}+p_{i\perp}~,
\label{z34}
\en
so that
\eqn
p^2_i = s\alpha_i\beta_i+p^2_{i\perp}.
\enn
In the region under consideration, i.e. for
\eq
s \gg |t| \sim m^2_A \sim m^2_B~,
\label{z35}
\en
the essential kinematics in Eq.(\ref{z32}) is defined by three conditions:
\eq
i)~~~~~~~~~~~~~~~~~~|p^2_{A_1\perp}|\sim |p^2_{B_1\perp}
|\sim |p^2_{G\perp}|\sim |t|~,
\label{z36}
\en
which means that all transverse momenta are limited;
\eq
ii)~~~~~~~\beta_{A_1}\sim \alpha_{B_1}
\sim 1~,~~~~~~~~\alpha_{A_1}\sim
\beta_{B_1}\sim \frac{|t|}{s}~,
\label{z37}
\en
which states that the order of magnitude of quark longitudinal momenta does
not change;
\eq
iii)~~~~~~~~~~~~\frac{|t|}{s}\la \beta_G \la 1~,~~~~~~~~\frac{|t|}{s}\la
\alpha_G \la 1~,
\label{z38}
\en
according to which the range of variation of the gluon rapidity is large. Since
\eqn
s\alpha_G\beta_G = -p^2_{G\perp} \sim |t|~,
\enn
only one independent variable varies strongly in the multi-Regge region,
leading to a contribution proportional to $\ln(s)$ \cite{FKL}.
\vskip.3cm
In LLA, where the discontinuity should be calculated with logarithmic
accuracy, only the multi-Regge kinematics could contribute so that the
region was defined by Eq.(\ref{z36}) and by
\eqn
\beta_{A_1} \approx 1~,~~~~~~~\alpha_{B_1} \approx 1~,~~~~~~~\alpha_{A_1}
\sim
\beta_{B_1} \sim \frac{|t|}{s}~,
\enn
\eq
\frac{|t|}{s}\ll \beta_G\ll 1~,~~~~~~~~~\frac{|t|}{s}\ll \alpha_G\ll 1~.
\label{z39}
\en
Now we need to keep a constant term together with the logarithmic one,
therefore, besides the contributions of the region (\ref{z38}) we must also
calculate those ones of the fragmentation region of the particles $A$, $A'$
and $B$, $B'$. Let us start with the first of them. The region is determined,
together with Eq.(\ref{z36}), by the relations
\eqn
\beta_{A_1} \approx 1~,~~~~~~~\alpha_{B_1} \approx 1~,~~~~~~~\alpha_{A_1}
\sim
\beta_{B_1} \sim \frac{|t|}{s}~;
\enn
\eq
\beta_G \sim 1~,~~~~~~~\alpha_G \sim \frac{|t|}{s}~.
\label{z40}
\en
Within the accuracy here required, the gauge invariant expression for the
amplitude ${\cal A}^{A_1GB_1}_{AB}$ in this region is given by
\eqn
{\cal A}^{A_1GB_1}_{AB} = \frac{2g^3}{s}\bar u(p_{A_1})\left
\{\langle A_1| T^jT^c| A \rangle\pnot_{+}\frac{(\pnot_A-\pnot_G+m_A)}
{-2(p_Ap_G)}\enot^*_G\right.
\enn
\eqn
\left. + \langle A_1|T^cT^j| A \rangle\enot^*_G\frac{(\pnot_{A_1}+\pnot_G+
m_A)}{2(p_{A_1}p_G)}\pnot_{+}-if_{ijc}\langle A_1|T^i| A \rangle\right.
\enn
\eqn
\times\left. \frac{2}{(p_{A_1}-p_A)^2}\left[\enot^*_G(p_{+}p_G)-
\pnot_G(p_{+}e^*_G)-
\pnot_{+}\left(\left(p_A-p_{A_1}+\frac{(p_B-p_{B_1})^2}{2p_Bp_G}p_B\right)
e^*_G\right)\right]\right\}
\enn
\eq
\times u(p_A)\epsilon^{*c}_G\langle B_1|T^j| B \rangle\frac{1}
{(p_{B_1}-p_B)^2}\bar u(p_{B_1})\pnot_{-}u(p_B)~.
\label{z41}
\en
All Feynman diagrams for the gluon emission in the quark-quark scattering
(see Fig. 2) contribute to the amplitude (\ref{z41}) and one may easily
recognize the contribution of each diagram in the Feynman gauge. Let
us remind that, if a diagram contains a gluon line connecting two parts
with strongly different Sudakov variables, then for calculating its asymptotic
contribution it is convenient to decompose $g^{\mu\nu}$ in the gluon
propagator in the form
\eq
g^{\mu\nu} = 2\frac{(p^{\mu}_{-}p^{\nu}_{+}+p^{\mu}_{+}p^{\nu}_{-})}{s}
+g^{\mu\nu}_\perp~.
\label{z42}
\en
The asymptotic contribution is given only by the first term in the RHS of
Eq.(\ref{z42}), supposing that the Sudakov variables of $\beta$-type
($\alpha$-type) at the $\mu$-vertex are much smaller (much larger) than
those at
the $\nu$-vertex. Clearly, this trick is used for the gluon lines with
momentum $q_2$ in the diagrams of Fig. 2(a,b,c). It is also used for the
gluon lines with momentum $q_1$ in the diagrams of Fig. 2(d,e), because
the dependence on the gluon momentum is factorized for these diagrams, in
the same way as for the soft gluon emission, and after this we have all the
conditions for applying the trick.
\vskip.3cm
We could perform all calculations in a gauge invariant manner, but it
occurs that a suitable choice of the gauge makes our task easier.
This happens in the axial gauge
\eq
e_Gp_{+} = 0~.
\label{z43}
\en
Using then this gauge condition, let us express the amplitude (\ref{z41}) in
terms of the Sudakov variables. First of all we introduce the simpler
notations
\eqn
k \equiv p_G~,~~~~~~~e \equiv e_G~,~~~~~~~q_1 \equiv p_A-p_{A_1}~,~~~~~~~
q_2 \equiv
p_{B_1}-p_B~,
\enn
\eq
k = q_1-q_2 = \alpha p_{+}+\beta p_{-}+k_{\perp}~.
\label{z44}
\en
Moreover, by means of the gauge condition (\ref{z43}) and of the transversality
requirement $ek=0$, we can express the polarization vector in terms of its
transverse part:
\eq
e = e_{\perp} - p_{+}\frac{e_{\perp}k_{\perp}}{p_{+}k}~.
\label{z45}
\en
Hence all the invariants necessary to calculate the amplitude
${\cal A}^{A_1GB_1}_{AB}$ acquire the following expression in terms of the
Sudakov variables:
\eqn
2p_{+}k = s\beta~,~~~~~~~2p_{-}k = -\frac{k^2_{\perp}}{\beta}~,~~~~~~~
2p_{+}p_{A_1} = s(1-\beta)~,
\enn
\eqn
2p_{-}p_{A_1} = \frac{m^2_A-q^2_{1\perp}}{1-\beta}~,~~~~~~~2p_Ak =
\frac{m^2_A{\beta}^2-{k_{\perp}}^2}{\beta}~,
\enn
\eq
2p_{A_1}k = \frac{m^2_A{\beta}^2-(k+{\beta}q_2)^2_\perp}{{\beta}(1-\beta)}~,
{}~~~~~~~q^2_1 = \frac{q^2_{1\perp}-m^2_A{\beta}^2}{1-\beta}~,~~~~~~~q^2_2 =
q^2_{2\perp}~.
\label{z46}
\en
Using now the polarization vector (\ref{z45}), the invariants (\ref{z46})
and the commutation relation (\ref{z18}) for the term of Eq.(\ref{z41})
containing $\langle A_1|T^cT^j| A \rangle$, through a simple calculation of
spinor algebra, we arrive at
\eqn
{\cal A}^{A_1GB_1}_{AB} = 2g^3\bar u(p_{A_1})\left\{\langle A_1|
T^jT^c| A \rangle[L(k_{\perp})-L(k_{\perp}+{\beta}q_{2\perp})]\right.
\enn
\eq
\left. -if_{ijc}\langle A_1|T^i| A \rangle[L(q_{1\perp})-L(k_
{\perp}+{\beta}q_{2\perp})]\right\}\pnot_{+}u(p_A)
\langle B_1|T^j| B \rangle\frac{1}{q^2_{2\perp}}\epsilon^{*c}_G
\delta_{\lambda_B,\lambda_{B_1}}~.
\label{z47}
\en
Here we have put
\eq
L(k_{\perp}) = L_{\mu}(k_{\perp})e^{*\mu}_{\perp}~,~~~~~L^{\mu}(k_{\perp}) =
\frac{\gamma^{\mu}_{\perp}(m_A{\beta}^2-{\beta}\knot_{\perp})+2k^{\mu}_
{\perp}}{m^2_A{\beta}^2-k^2_{\perp}}~.
\label{z48}
\en
In order to go from Eq.(\ref{z41}) to Eq.(\ref{z47}) we must take into
account that in the helicity basis one has
\eq
\bar u(p_{B_1})\pnot_{-}u(p_B) = s\delta_{\lambda_B,\lambda_{B_1}}~.
\label{z49}
\en
The corresponding expression for the amplitude ${\cal A}^{A_1GB_1}_{A'B'}$
can be obtained in the same way and reads
\eqn
{\cal A}^{A_1GB_1}_{A'B'} = 2g^3\bar u(p_{A_1})\left\{\langle A_1|T^jT^c|
A' \rangle(L(k_{\perp}+{\beta}q_{\perp})-L(k_{\perp}+{\beta}q_{2\perp}))
\right.
\enn
\eqn
\left. -if_{ijc}\langle A_1|T^i| A' \rangle(L(q_{1\perp}-
(1-{\beta})q_{\perp})-L(k_{\perp}+{\beta}q_{2\perp}))\right\}
\enn
\eq
\times\pnot_{+}
u(p_{A'})\epsilon^{*c}_G\frac{1}{(q_2-q)^2_{\perp}}\langle B_1|T^j| B' \rangle
\delta_{\lambda_{B'},\lambda_{B_1}}~.
\label{z50}
\en
Let us note that this expression follows from Eq.(\ref{z47}) by simple
substitutions. Indeed, the amplitude ${\cal A}^{A_1GB_1}_{A'B'}$ must take
the same form of ${\cal A}^{A_1GB_1}_{AB}$ when
expressed in terms of the {\it primed} Sudakov variables and {\it primed}
vectors $p'_{\pm}$, defined by
\eqn
p_{A'} = p^{'}_{-}+\frac{m^2_A}{s}p{'}_{+}~,~~~~~~~~~p_{B'} =
p^{'}_{+}+\frac{m^2_B}{s}p{'}_{-}~,
\enn
\eq
p_i = \beta^{'}_ip^{'}_{-}+{\alpha^{'}}_ip^{'}_{+}+p_{i\perp^{'}}~,
\label{z51}
\en
where $\perp^{'}$ means transversal to the $(p_{A'},p_{B'})$ plane.
Since
\eqn
p_{A'} = p_{-}+\frac{m^2_A-q^2_{\perp}}{s}p_{+}-q_{\perp}~,
\enn
\eq
p_B{'} = p_{+}+\frac{m^2_B-q^2_{\perp}}{s}p_{-}+q_{\perp}~,
\label{z52}
\en
we get the following connection between the basis vectors:
\eqn
p'_{+}\simeq p_{+}-\frac{q^2_{\perp}}{s}p_{-}+q_{\perp}~,
\enn
\eq
p'_{-}\simeq p_{-}-\frac{q^2_{\perp}}{s}p_{+}-q_{\perp}~,
\label{z53}
\en
and between the Sudakov variables:
\eqn
\beta^{'}_i \simeq \beta_i-\frac{q^2_{\perp}}{s}\alpha_i+2\frac{q_{\perp}
p_{i\perp}}{s}~,~~~~\alpha^{'}_i \simeq \alpha_i-\frac{q^2_{\perp}}{s}\beta_i
-2\frac{q_{\perp}p_{i\perp}}{s}~,
\enn
\eq
p_{i\perp'} \simeq  p_{i\perp}+(\beta_i-\alpha_i)q_{\perp}+2\frac{p_{+}-
p_{-}}{s}\left[(\beta_i-\alpha_i)q^2_{\perp}+q_{\perp}p_{i\perp}\right]~.
\label{z54}
\en
Keeping only the leading terms, in the kinematical region defined by Eqs.
(\ref{z36}), (\ref{z37}) and (\ref{z40}), we find that the amplitude
${\cal A}^{A_1GB_1}_{A'B'}$ can be obtained from the RHS of Eq.(\ref{z47}) by
the substitution
\eq
k_{\perp} \rightarrow k_{\perp}+\beta{q_{\perp}}~,~~~~~~~~q_{2\perp} =
p_{B_{1}\perp} \rightarrow q_{2\perp}-q_{\perp}~,
\label{z55}
\en
besides the obvious change $A \rightarrow A'$ and $B \rightarrow B'$.
\vskip.3cm
We are now able to calculate the discontinuity (\ref{z32}). Using Eqs.
(\ref{z47}) and (\ref{z50}) and performing the summation over colour and spin
states of the intermediate particles, we obtain
\eqn
\sum_{{A_1},G,{B_1}}{\cal A}^{A_1GB_1}_{AB}{\cal A}^{\ast A_1GB_1}_{A'B'} =
\frac{4g^6\delta_{{\lambda_B},\lambda_{B'}}}
{q^2_{2\perp}(q_2-q)^2_{\perp}}\left[\langle A^{'}|T^cT^{j^{'}}
T^jT^c| A \rangle R_a\right.
\enn
\eqn
\left. +f_{i^{'}j^{'}c}f_{ijc}\langle A^{'}|T^{i^{'}}T^i| A \rangle R_{na}
+if_{ij^{'}c}\langle A^{'}|T^iT^jT^c| A \rangle R_{int}\right.
\enn
\eq
\left. -if_{ijc}\langle A^{'}|T^cT^{j^{'}}T^i| A \rangle R^{'}_{int}\right]
\langle B'|T^{j^{'}}T^j| B \rangle~.
\label{z56}
\en
Here the first two terms in the square brackets come respectively from the
product of the {\it abelian} and {\it non abelian} parts of the amplitudes
(\ref{z47}) and (\ref{z50}); one has
\eqn
R_a = \bar u(p_{A'})\pnot_{+}\left[\bar L^{\mu}(k_{\perp}+\beta q_
{2\perp})-\bar L^{\mu}(k_{\perp}+\beta q_{\perp})\right]\left(\pnot_{A_1}
+m_A\right)
\enn
\eq
\times \left[L_{\mu}(k_{\perp})-L_{\mu}(k_{\perp}+\beta q_{2\perp})\right]
\pnot_{+}u(p_A)~,
\label{z57}
\en
with
\eq
\bar L^\mu(k_{\perp}) = \frac{(m_A{\beta}^2-{\beta}\knot_{\perp})\gamma^{\mu}_
{\perp}+2k^{\mu}_{\perp}}{m^2_A{\beta}^2-k^2_{\perp}}
\label{z58}
\en
and
\eqn
R_{na} = \bar u(p(_{A'})\pnot_{+}\left[\bar L^{\mu}(k_{\perp}+\beta q_
{2\perp})-\bar L^{\mu}(q_{1\perp}-(1-{\beta})q_{\perp})\right]
\left(\pnot_{A_1}+m_A\right)
\enn
\eq
\times \left[L_{\mu}(q_{1\perp})-L_{\mu}(k_{\perp}+\beta q_{2\perp})
\right]\pnot_{+}u(p_A)~.
\label{z59}
\en
The remaining terms come from the interference of the {\it abelian} and
{\it non abelian} parts. It is not necessary to write them explicitely,
because they do not contribute to the discontinuity which we
are interested in. Remind that we need to calculate the helicity conserving
part of the discontinuity for the amplitude with colour octet state in the
$t$ channel and negative signature. As it was discussed just before writing
the relations (\ref{z18}), we must antisymmetrize the colour
factor $\langle B'|T^{j^{'}}T^j| B \rangle$ with respect to the inversion
of the order of the group generators (antisymmetrization of the colour
factor for the particle $A$ leads to the same result).
\vskip.3cm
Bearing this in mind, we first calculate the colour factors
in Eq.(\ref{z56}). As for the $R_{na}$ term,
using the relations (\ref{z18}) and taking into account that the second of them
gives us (in the case of adjoint representation $(T^i)_{lm}=-if^i~_{lm}$)
\eq
f_{ijk}f_{ilm}f_{jmn} = \frac{N}{2}f_{knl}~,
\label{z60}
\en
we get
\eqn
\frac{1}{2}\langle B'|T^k| B \rangle if_{kj^{'}j}f_{i^{'}j^{'}c}f_{ijc}
\langle A'|T^{i^{'}}T^i| A \rangle =
\enn
\eqn
\frac{N}{4}\langle B'|T^k| B \rangle if_{ki^{'}i}\langle A'|T^{i^{'}}T^i|
A \rangle =
\enn
\eq
-\frac{N^2}{8}\langle B'|T^k| B \rangle\langle A'|T^k| A \rangle~.
\label{z61}
\en
The same relations allow to show that the colour factors for the
interference terms are zero. For example, the colour factor of the $R_{int}$
term is
\eqn
\frac{1}{2}if_{kj^{'}j}\langle B'|T^k| B \rangle if_{ij^{'}c}\langle A'
|T^iT^jT^c| A \rangle =
\enn
\eqn
\frac{1}{2}if_{kj^{'}j}\langle B'|T^k| B \rangle if_{ij^{'}c}\left(if_{ljc}
\langle A'|T^iT^l| A \rangle + \langle A'|T^iT^cT^j| A \rangle \right) =
\enn
\eq
\frac{1}{2}\langle B'|T^k| B \rangle \left(\frac{N}{2}if_{kli}
\langle A'|T^iT^l| A \rangle +\frac{N}{2}if_{kj^{'}j}\langle A'|T^{j^{'}}T^j
| A \rangle\right) = 0~.
\label{z62}
\en
The same result is obtained for $R^{'}_{int}$. For the
colour factor of the $R_a$
term we find
\eqn
\frac{i}{2}f_{kj^{'}j}\langle B'|T^k| B \rangle
\langle A^{'}|T^cT^{j^{'}}T^jT^c| A\rangle =
\enn
\eq
-\frac{N}{4}\langle B'|T^k| B \rangle\langle A^{'}|T^cT^kT^c| A\rangle =
\frac{1}{8}\langle B'|T^k| B \rangle\langle A^{'}|T^k| A\rangle~.
\label{z63}
\en
To obtain this result we used the relation
\eq
T^aT^cT^a = \left(C_2-\frac{N}{2}\right)T^c~.
\label{z64}
\en
Here $C_2$ is the eigenvalue of the Casimir operator $T^aT^a$ whose
expression for quarks is
\eqn
C_2 = \frac{N^2-1}{2N}~.
\enn
In turn eq.(\ref{z64}) can be obtained using the relations (\ref{z18}).
\vskip 0.3cm
It remains now to calculate $R_a$ and $R_{na}$. It is clear from
Eqs.(\ref{z57}) and (\ref{z59}) that both quantities can be obtained from
the general expression
\eq
G(r_{\perp},r'_{\perp}) = \bar u(p_{A'})\pnot_{+}\bar L^{\mu}(r_{\perp})
\left(\pnot_{A_1}+m_A\right)L_{\mu}(r_{\perp}+r'_{\perp})\pnot_{+}u(p_A)~,
\label{z65}
\en
where $r_{\perp}$ is linear in $k_{\perp}$ and $r'_{\perp}$ does not depend
on $k_{\perp}$. Taking into account that
\eqn
p_{+}{\gamma}_{\perp} = 0~,~~~~~\pnot^{~2}_{+} = 0~,~~~~~2p_{+}p_{A_1} =
s(1-\beta)~,
\enn
and putting
\eq
d(l)={1\over m^2_{A}\beta^2-l^2}~,
\label{zqa1}
\en
we find that the expression (\ref{z65}) may be written as
\eqn
G(r_{\perp},r'_{\perp}) = s(1-\beta)d(r_\perp)
d(r_\perp+r'_\perp)\bar u(p_{A'})
\enn
\eqn
\times \left[\left(-m_A{\beta}^2-\beta\rnot_{\perp}\right)
{\gamma}^{\mu}_{\perp}+2r^{\mu}_{\perp}\right]\left[{\gamma}_{{\perp}\mu}
\left(m_A{\beta}^2-\beta(\rnot+\r'not)_{\perp}\right)+2(r+r')_{{\perp}{\mu}}
\right]\pnot_{+}u(p_A) =
\enn
\eqn
s(1-\beta)d(r_\perp)d(r_\perp+r'_\perp)
\bar u(p_{A'})\left[(D-2)\left(-m^2_A{\beta}^4+{\beta}^2
r_\perp\left(r+r'\right)_{\perp}
+m_A{\beta}^3\r'not_
{\perp}\right)\right.
\enn
\eq
\left. +4r_{\perp}(r+r')_{\perp}(1-\beta)-2m_A{\beta}^2\r'not_
{\perp}\right]\pnot_{+}u(p_A)~.
\label{z66}
\en
We have used here the fact that the term $\rnot_{\perp}\r'not_{\perp}$, being
under integration over azimutal angles, is equivalent to
$r_{\perp}r'_{\perp}$.
\vskip.3cm
The term $\r'not_{\perp}$, after the integration is
performed, becomes proportional to $\qnot_{\perp}$, therefore it can be
omitted. In fact, because of the relation (see, for instance, Eq.(37)
in Ref.~\cite{FFQ1})
\eqn
\bar u(p_{A'})\qnot_{\perp}\pnot_{+}u(p_A) = \bar u(p_{A'})(\pnot_A-
\pnot_{A'})\pnot_{+}u(p_A) =
\enn
\eq
\bar u(p_{A'})(-2m\pnot_{+}+s)u(p_A) = s\left(\bar u(p_{A'})u(p_A)-2m\delta_
{\lambda_{A},\lambda_{A'}}\right) = -is\sqrt{-t}\delta_{\lambda_{A},-
\lambda_{A'}}~,
\label{z67}
\en
it does not contribute to the helicity conserving part which we are
interested in. After some algebra, for this part we obtain
\eqn
G^{(+)}(r_{\perp},r'_{\perp})=s^2(1-\beta)
d(r_\perp)d(r_\perp+r'_\perp)\left[-\left(2(1-\beta)+(D-2)
\frac{{\beta}^2}{2}\right)\right.
\enn
\eq
\left. \left(r'^2_{\perp}+{1\over d(r_\perp+r'_\perp)}
+{1\over d(r_\perp)}\right)+4m^2_A{\beta}^2(1-\beta)\right]~,
\label{z68}
\en
where the superscript $(+)$ stands for helicity conserving part.\par
Let us denote by $\Delta^{(3A)}$ the contribution to
$\Delta^{(3)}_S$ defined by Eqs.(\ref{z11}) and (\ref{z13}),
which comes from the
fragmentation region of the particles $A$ and $A'$. Applying
the unitarity condition (\ref{z32}) and Eq.(\ref{z56}) with the colour
factors given by formulas (\ref{z61})-(\ref{z63}), we have
\eq
\Delta^{(3A)} = \frac{4g^4t}{s}\int\frac{d\Phi_3\left(p_A+p_B;p_{A_1},k,
p_{B_1}\right)}{{2\pi}q^2_{2\perp}(q_2-q)^2_{\perp}}\left[-\frac{1}{8}
R^{(+)}_a+\frac{N^2}{8}R^{(+)}_{na}\right]~
\label{z69}~,
\en
where the superscript $(+)$ means, as before, helicity conserving part. The
integration runs over the fragmentation region of the particles $A$ and
$A'$, where the phase space element takes the form
\eq
\frac{d\Phi_3\left(p_A+p_B;p_{A_1},k,p_{B_1}\right)}{2\pi} =
\frac{1}{4s}\frac{d{\beta}}{\beta(1-\beta)}\frac{d^{(D-2)}q_{1\perp}d^
{(D-2)}q_{2\perp}}{(2\pi)^{2(D-1)}}~.
\label{z70}
\en
Here
\eqn
q_1 = p_A-p_{A_1}~,~~~~~~~q_2 = p_{B_1}-p_B~,~~~~~~~k = q_1-q_2~,
\enn
\eqn
{\beta}_0 \leq \beta \leq 1~,
\enn
${\beta}_0$ being a not specified, artificially introduced boundary of
the fragmentation region of the particles $A$ and $A'$. It is convenient to
split $\Delta^{(3A)}$ into two parts,
\eq
\Delta^{(3A)} = \Delta^{(3A)}_a+\Delta^{(3A)}_{na}~,
\label{z71}
\en
which respectively contain the terms $R^{(+)}_a$ and $R^{(+)}_{na}$. Using
the phase space element (\ref{z70}), the expressions (\ref{z57}) and
(\ref{z59}) for $R^{(+)}_a$ and $R^{(+)}_{na}$ respectively and
Eqs.(\ref{z65}), (\ref{z68}) for $G(r_{\perp},r'_{\perp})$,
from Eq.(\ref{z69}) (here
and below all vectors are ($D-2$)-dimensional and orthogonal to the
$(p_A,p_B)$ plane) we get
\eqn
\Delta^{(3A)}_a = \frac{g^4t}{8}\int\frac{d^{(D-2)}q_1}{(2\pi)^{D-1}}
\frac{d^{(D-2)}q_2}{(2\pi)^{D-1}}\int^1_{{\beta}_0}\frac{d{\beta}}{\beta}
\frac{{\beta}^2}{q^2_2(q_2-q)^2}\left[\left(2(1-{\beta})+\frac{{\beta}^2}
{2}(D-2)\right)\right.
\enn
\eqn
\left. \times d(k)d(k+\beta q)d(k+\beta q_2)
\left({-q^2\over d(k+\beta q_2)}+{q^2_2\over
d(k+\beta q)}+{(q_2-q)^2\over d(k)}\right)\right.
\enn
\eq
\left.+4m^2_A(1-{\beta})\left(d(k+\beta q_2)-
d(k)\right)\left(d(k+\beta q_2)
-d(k+\beta q)\right)\vbox to 19.08pt{}\right],
\label{z72}
\en
and
\eqn
\Delta^{(3A)}_{na} = \frac{g^4N^2t}{8}\int\frac{d^{(D-2)}q_1}{(2\pi)^{D-1}}
\frac{d^{(D-2)}q_2}{(2\pi)^{D-1}}\int^1_{{\beta}_0}\frac{d{\beta}}{\beta}
\frac{(1-{\beta})^2}{q^2_2(q_2-q)^2}\left[\left(2(1-{\beta})+
\frac{{\beta}^2}
{2}(D-2)\right)\right.
\enn
\eqn
\left. \times d(q_1)d(k+\beta q_2)d(q_1-(1-\beta)q)
\left({q^2\over d(k+\beta q_2)}-{q^2_2\over d(q_1-(1-\beta)q)}-
{(q_2-q)^2\over d(q_1)}\right)\right.
\enn
\eq
\left.-\frac{4m^2_A\beta^2}
{1-{\beta}}\left(d(k+\beta q_2)-d(q_1)\right)
\left(d(k+\beta q_2)-d(q_1-(1-\beta)q)\right)\right]~.
\label{z73}
\en
\vskip.3cm
Eqs.(\ref{z71})-(\ref{z73}) give us the contribution of the fragmentation
region of the particles $A$ and $A'$. It should be supplemented by two
contributions. One of them comes from the fragmentation region of the particles
$B$ and $B'$ which, in analogy to Eq.(\ref{z71}), can be divided into two
parts accordingly to the colour factors,
\eq
\Delta^{(3B)} = \Delta^{(3B)}_a+\Delta^{(3B)}_{na}~.
\label{z75}
\en
Evidently, $\Delta^{(3B)}_a$ and $\Delta^{(3B)}_{na}$ can be respectively
obtained from Eqs.(\ref{z72}) and (\ref{z73}) by the substitutions
\eqn
q_{1,2} \rightarrow -q_{2,1}~,~~~~~~~q \rightarrow -q~,~~~~~~~
m_A \rightarrow m_B~,
\enn
\eq
\beta \rightarrow \alpha \equiv {\alpha}_G = \frac{-k^2}{s\beta}~,~~~~~~~
{\beta}_0 \rightarrow {\alpha}_0~.
\label{z76}
\en
The other contribution, that we call $\Delta^{(3I)}$, comes from the
multi-Regge region, defined by Eqs.(\ref{z36}) and
(\ref{z39}), which is
intermediate between the two fragmentation regions. This contribution was
calculated \cite{FKL} and reads
\eqn
\Delta^{(3I)} = \frac{g^4N^2t}{8}\int\frac{d^{(D-2)}q_1}{(2\pi)^{D-1}}
\frac{d^{(D-2)}q_2}{(2\pi)^{D-1}}\int^{{\beta}_0}_{-\frac{k^2}{s\alpha_0}}
\frac{d{\beta}}{\beta}\frac{1}{q^2_2(q_2-q)^2}
\enn
\eq
\times 2\left[\frac{q^2}{q^2_1(q_1-q)^2}-\frac{q^2_2}{q^2_1k^2}-
\frac{(q_2-q)^2}{(q_1-q)^2k^2}\right]~.
\label{z77}
\en
This result can be obtained from the expression
(\ref{z73}) for $\Delta^{(3A)}_{na}$ by changing the limits of integration
over $\beta$ and going to small $\beta$ in the integrand.
Notice that the abelian part (\ref{z72}) does not contribute in this case.
It means that the integrands of Eqs.(\ref{z72}) and (\ref{z73}) are valid
in a region wider than that of the fragmentation one, namely not simply
for $\beta \sim 1$ but for $|t|/s\ll\beta \leq 1$. This is not
unusual because $|t|/s$ is not an artificial bound but a natural
one where the applicability should be broken. That allows us not to
consider the multi-Regge region separately, but to include it in any
of the two fragmentation regions
which, then,
become overlapping. Hence, the total three particle discontinuity can be
obtained as the sum of the contributions of the two fragmentation regions,
\eq
\Delta^{(3)}_{S} = \Delta^{(3A)}+\Delta^{(3B)}~.
\label{z78}
\en
Here the first term in the RHS is given by Eqs.(\ref{z71})-(\ref{z73}),
the other one follows from it by the substitution (\ref{z76}), with the
parameters ${\beta}_0$ and ${\alpha}_0$ satisfying the condition
$s{\beta}_0{\alpha}_0=-k^2$. Choosing
\eq
{\beta}_0 = {\alpha}_0 = \sqrt{\frac{-k^2}{s}}~,
\label{z79}
\en
we get $\Delta^{(3B)}$ from $\Delta^{(3A)}$ simply by
the substitution $m_A \rightarrow m_B$.
\vskip.3cm
As for the abelian part $\Delta^{(3A)}_{a}$ of the last discontinuity, it is
clear from Eq.(\ref{z72}) that ${\beta}_0$ could be put equal to zero
because the small $\beta$ region does not contribute. This is in
accordance with our experience in QED, where it is well known that cones of
photon emission by two scattered particles do not overlap at high energies.
\vskip.3cm
Performing the integration over $q_1$ in Eq.(\ref{z72}) with the help
of the Feynman parametrization,
\eq
\int\frac{d^{(D-2)}r}{ (2{\pi})^{D-1}}d(r)d(r+l) =
\frac{2\Gamma\left(3-\frac{D}{2}\right)}{(4\pi)^{\frac{D}{2}}}\int^1_0
\frac{dx}{c(m_A\beta,l)}~,
\label{z80}
\en
where
\eq
c(m,l) = [m^2-l^2x(1-x)]^{3-\frac{D}{2}}~,
\label{zqa2}
\en
we find
\eqn
\Delta^{(3A)}_a = \frac{g^4t}{4}\frac{\Gamma\left(3-\frac{D}{2}\right)}
{(4\pi)^{\frac{D}{2}}}\int\frac{d^{(D-2)}q_2}{(2{\pi})^{D-1}q^2_2
\left(q_2-q\right)^2}
\enn
\eqn
\times\int^1_0\frac{d{\beta}}{\beta^{(5-D)}}
\int^1_0dx
\left[\left(2(1-{\beta})+\frac{{\beta}^2}{2}(D-2)\right)\right.
\enn
\eq
\left. \times\left({-q^2\over c(m_A,q)}
+{2q^2_2 \over c(m_A,q_2)}\right)
+4m^2_A(1-{\beta})\left({1\over c(m_A,0)}+{1\over c(m_A,q)}
-{2\over c(m_A,q_2)}\right)\right]~.
\label{z81}
\en
After this step the integration over ${\beta}$ can be easily performed and
yields
\eqn
\Delta^{(3A)}_{a} = \frac{g^4t}{4}\frac{\Gamma\left(2-\frac{D}{2}\right)}
{(4\pi)^{\frac{D}{2}}}
\int\frac{d^{(D-2)}q'}{(2{\pi})^{D-1}q'^{~2}\left(q'-q\right)^2}
\enn
\eqn
\times\int^1_0dx\left[\left({D-4\over 4}+{1\over D-3}\right)
\left({q^2\over c(m_A,q)}-{2q'^{~2}\over c(m_A,q')}\right)\right.
\enn
\eq
\left.+{2m_A^2\over D-3}
\left({2\over c(m_A,q')}-{1\over c(m_A,q)}-{1\over c(m_A,0)}\right)\right]~.
\label{z82}
\en
\vskip.3cm
In order to calculate the non abelian contribution $\Delta^{(3A)}_{na}$
given by Eq.(\ref{z73}) with ${\beta}_0=\sqrt{-k^2/s}$, we
apply the following trick: we subtract and add this contribution considered
for a massless quark, $\Delta^{(3A)}_{na}(m_A=0)$. While performing the
integration over $q_1$ of the difference
\eq
\delta^{(3A)}_{na} = \Delta^{(3A)}_{na}-\Delta^{(3A)}_{na}(m_A=0)~,
\label{z83}
\en
we may put ${\beta}_0=0$. With the help of Eq.(\ref{z80}) we arrive at
\eqn
\delta^{(3A)}_{na} = \frac{g^4N^2t}{4}\frac{ \Gamma\left(3-\frac{D}{2}\right)}
{(4\pi)^{\frac{D}{2}}}
\int\frac{d^{(D-2)}q_2}{(2{\pi})^{D-1}q^2_2
\left(q_2-q\right)^2}\int^1_0d{\beta}\frac{(1-{\beta})^2}{\beta}
\enn
\eqn
\times\int^1_0dx\left\{\left(2(1-{\beta})+
\frac{{\beta}^2}{2}(D-2)\right)
\left[q^2\left(\frac{1}{c(m_A\beta,q(1-\beta))}-\frac{1}
{c(0,q(1-\beta))}\right)\right.\right.
\enn
\eqn
-2q^2_2\left.\left.
\left(\frac{1}{c(m_A\beta,q_2(1-\beta))}-\frac{1}{c(0,q_2(1-\beta))}
\right)\right]\right.
\enn
\eq
\left. -\frac{4m^2_A{\beta}^2}{1-\beta}\left[\frac{1}
{c(m_A\beta,q(1-\beta))}
-\frac{2}
{c(m_A\beta,q_2(1-\beta))}+\frac{1}{c(m_A\beta,0)}\right]\right\}~,
\label{z84}
\en
The total contribution of the three particle intermediate state to the
discontinuity is then given by the sum
\eq
\Delta^{(3)}_S = \Delta^{(3A)}_{a}+\delta^{(3A)}_{na}+\Delta^{(3B)}_{a}+
\delta^{(3B)}_{na}+\Delta^{(3A)}_{na}(m_A=0)+\Delta^{(3B)}_{na}(m_B=0)~.
\label{z86}
\en
Here $\Delta^{(3A)}_{a}$ and $\delta^{(3A)}_{na}$ are given by Eqs.
(\ref{z82}) and (\ref{z84}) respectively, $\Delta^{(3B)}_{a}$ and
$\delta^{(3B)}_{na}$ follow from these equations by the substitution
$m_A \rightarrow m_B$ and $\Delta^{(3A)}_{na}(m_A=0)$, which is equal to
$\Delta^{(3B)}_{na}(m_B=0)$, comes from Eq.(\ref{z73}) with
${\beta}_0=\sqrt{-k^2/s}$.

\vskip 0.3cm

{\bf 5. TWO-LOOP CORRECTION TO THE GLUON TRAJECTORY}

We are now ready to discuss the formula (\ref{z12}) for the correction
$\omega^{(2)}(t)$. It contains the
discontinuity $\Delta_S$ which has been calculated in this paper and,
according to Eq.(\ref{z13}), is given by the sum of the contributions
$\Delta^{(2)}_S$ and $\Delta^{(3)}_S$, respectively quoted in
Eqs.(\ref{z21}) and (\ref{z86}). Besides that, Eq.(\ref{z12}) involves
the leading
contribution to the trajectory $\omega^{(1)}(t)$ shown in Eq.(\ref{z4}) and
the one-loop corrections to the $QQR$ vertex $\Gamma^{(+)}_{QQ}$, which are
given by Eqs.(\ref{z22})-(\ref{z24}), (\ref{z27}), (\ref{z28})
and (\ref{z31}).
Notice that the discontinuity $\Delta^{(2)}_S$ itself is expressed in
terms of $\omega^{(1)}(t)$ and the $QQR$ vertex. Therefore, the correction
$\omega^{(2)}(t)$ looks rather complicated. Fortunately, a series of
remarkable cancellations occurs among various terms in it.
\vskip.3cm
First of all, the term $\Delta^{(3A)}_a$ in $\Delta^{(3)}_S$ cancels all the
terms in Eq.(\ref{z12}) which contain the contribution $a_Q(r,m^2_A)$ (see
Eq.(\ref{z24})) to the $QQR$ vertex $\Gamma^{(+)}_{AA}$ defined in
Eq.(\ref{z22}). It is worthwhile to note here that this vertex enters
Eq.(\ref{z12}) for $\omega^{(2)}(t)$ not only explicitely, but also through
the two-particle discontinuity $\Delta^{(2)}_S$ (see Eq.(\ref{z13})).
One might easily see the cancellation putting the expressions (\ref{z24}) for
$a_Q(r,m^2_A)$ and (\ref{z4}) for $\omega^{(1)}(t)$ into Eqs.(\ref{z21}) and
(\ref{z12}), and comparing the result with Eq.(\ref{z82}). Of course, the same
cancellation exists between the term $\Delta^{(3B)}_a$ and the terms which
contain $a_Q(r,m^2_B)$. Quite analogously, the term $\delta^{(3A)}_{na}$
(\ref{z84}) in $\Delta^{(3)}_S$ cancels all the terms in Eq.(\ref{z12}) which
contain $\delta_g(r,m^2_A)$ (\ref{z31}). In order to make this
cancellation more evident, one only needs to change the variables in
Eq.(\ref{z31}):
\eq
x_1 \rightarrow \beta~,~~~~~~~~~~x_2 \rightarrow (1-\beta)x~.
\label{z87}
\en
In turn also the term $\delta^{(3B)}_{na}$ cancels the terms with
$\delta_g(r,m^2_B)$.
\vskip.3cm
After these cancellations are performed, one realizes that the correction
$\omega^{(2)}(t)$ of Eq.(\ref{z12}) does not depend on the masses of the
scattered quarks. This independence was expected because the trajectory, by
definition, cannot depend on the properties of the scattered particles.
In fact, Eq.(\ref{z12}), because of Eqs.(\ref{z86}), (\ref{z21}), (\ref{z22})
and (\ref{z27}), gives us
\eqn
\omega^{(2)}(t) = 2\Delta^{(3A)}_{na}(m_A=0)+\frac{g^2Nt}{(2\pi)^{D-1}}
\int\frac{d^{(D-2)}q_1}{q^2_1\left(q_1-q\right)^2}\left[\omega^{(1)}(q^2_1)
\ln\left(\frac{s}{-q^2_1}\right)\right.
\enn
\eq
\left. +2a_f(q^2_1)+2a_g(q^2_1,0)\vbox to 19.08pt{}\right]
-\left({\omega}^{(1)}(t)\right)^2\ln \left(\frac{s}{-t}\right)-2
{\omega}^{(1)}(t)\left[a_f(t)+a_g(t,0)\right]~,
\label{z88}
\en
where ${\omega}^{(1)}(t)$, $a_f(t)$, $a_g(t,0)$ and $\Delta^{(3A)}_{na}$
are given by Eqs.(\ref{z4}), (\ref{z23}), (\ref{z28}) and (\ref{z73})
respectively.
\vskip.3cm
Looking at Eq.(\ref{z88}), we observe that not all the cancellations are
performed up to this point. Indeed, the trajectory cannot depend on $s$,
therefore the terms with $\ln(s)$ must cancel each other. Let us note that the
cancellation of these terms is a consequence of the gluon Reggeization in
LLA, which was previously proved \cite{FKL,BLF}, and can serve as a
check of our calculations, whereas the cancellations discussed above
confirm the gluon Reggeization beyond LLA.
\vskip.3cm
In order to explicitely demonstrate the cancellation of the terms with
$\ln(s)$ and
to present the correction $\omega^{(2)}(t)$ in a more transparent form, we
integrate over $\beta$ in Eq.(\ref{z73}) with $m_A=0$. We go through this
step in the
following way. We divide the integration region into two parts, one from
${\beta}_0=\sqrt{-k^2/s}$ to $\delta$, the other one from $\delta$
to $1$, with $\delta \ll 1$ (at the end of the calculation we let
$\delta$ go to zero). In the first region we may put ${\beta}=0$ everywhere
besides the factor $d{\beta}/{\beta}$; thus the corresponding
contribution to $\Delta^{(3A)}_{na}(m_A=0)$ is given by
\eq
\frac{g^4N^2t}{8}\int\frac{d^{(D-2)}q_1}{(2\pi)^{D-1}}\frac{d^{(D-2)}q_2}
{(2\pi)^{D-1}}\frac{1}{q^2_1(q_2-q)^2}\left[\frac{q^2}{q^2_2(q_1-q)^2}-
\frac{2}{(q_1-q_2)^2}\right]\ln\left(\frac{s{\delta}^2}{-(q_1-q_2)^2}\right)~.
\label{z89}
\en
In the second region the change of the variable
$q_1 \rightarrow (1-{\beta})q_1$ factorizes the $\beta$-dependence, yielding
\eq
\int^1_{\delta}\frac{d{\beta}}{\beta}(1-{\beta})^{D-4}\left(2(1-{\beta})+
\frac{{\beta}^2}{2}(D-2)\right) =
2\ln\left(\frac{1}{\delta}\right)+2\psi(1)-2\psi(D-3)-\frac{3}{2(D-3)}~.
\label{z90}
\en
To obtain this result we have used the decomposition
\eq
\int^1_{\delta}\frac{d{\beta}}{\beta}f({\beta}) = \int^1_{\delta}
\frac{d{\beta}}{\beta}\left(f({\beta})-f(0)\right)+\int^1_{\delta}
\frac{d{\beta}}{\beta}f(0)~,
\label{z91}
\en
and the integral
\eq
\int^1_0\frac{d{\beta}}{\beta}\left[(1-{\beta})^{D-4}-1\right] =
\psi(1)-\psi(D-3)~.
\label{z92}
\en
Given Eqs.(\ref{z89}) and (\ref{z90}), from Eq.(\ref{z73}) we get
\eqn
\Delta^{(3A)}_{na}(m_A=0) = \frac{g^4N^2t}{8}\int\frac{d^{(D-2)}q_1}{(2\pi)^
{D-1}}\frac{d^{(D-2)}q_2}{(2\pi)^{D-1}}\frac{1}{q^2_1(q_2-q)^2}\left[\frac{q^2}
{q^2_2(q_1-q)^2}-\frac{2}{(q_1-q_2)^2}\right]
\enn
\eq
\times\left[\ln\left(\frac{s}{-(q_1-q_2)^2}\right)+2\psi(1)-2\psi(D-3)
-\frac{3}
{2(D-3)}\right]~.
\label{z93}
\en
We now substitute this expression, together with the expressions (\ref{z4})
for ${\omega}^{(1)}(t)$, (\ref{z23}) and (\ref{z28}) for $a_f(t)$
and $a_g(t,0)$ respectively, in Eq.(\ref{z88}). Making use of the equality
\eq
\int\frac{d^{(D-2)}r}{(2\pi)^{D-1}r^2(r-r')^2} = -\frac{4\Gamma
\left(2-\frac{D}{2}\right)\Gamma^2\left(\frac{D}{2}-1\right)}
{(4\pi)^{\frac{D}{2}}\Gamma(D-3)\left(-r'^2\right)^{3-\frac{D}{2}}}
\label{z94}
\en
in order to represent the contribution of $a_g(t,0)$ in integral form, we
obtain, as final result,
\eqn
\omega^{(2)}(t) = \frac{g^4N^2t}{4}\int\frac{d^{(D-2)}q_1}{(2\pi)^{D-1}q^2_1}
\left\{\vbox to 22.9pt{}
\int\frac{d^{(D-2)}q_2}{(2\pi)^{D-1}q^2_2}\left[\frac{q^2}{(q_1-q)^2
(q_2-q)^2}\ln\left(\frac{q^2}{(q_1-q_2)^2}\right)\right.\right.
\enn
\eqn
\left.\left. +\frac{2}{(q_1+q_2-q)^2}\ln\left(\frac{q^2_1}{(q_1-q)^2}\right)+
\left(-\frac{q^2}{(q_1-q)^2(q_2-q)^2}+\frac{2}{(q_1+q_2-q)^2}
\right)\right.\right.
\enn
\eqn
\left.\left. \times\left(2\psi(D-3)+\psi{\left(3-\frac{D}{2}\right)}-
2\psi{\left(\frac{D}{2}-2\right)}-\psi(1)\right.\right.\right.
\enn
\eqn
\left.\left.\left. +\frac{1}{(D-3)}\left(\frac{1}{4(D-1)}-\frac{2}
{D-4}-\frac{1}{4}\right)\right)\right]-\frac{8}{N}\frac{ \Gamma\left(2-
\frac{D}{2}\right) }{(4\pi)^{\frac{D}{2}} }\frac{1}{(q_1-q)^2}\right.
\enn
\eq
\left. \times\sum_f\int_0^1 dxx(1-x)\left[\frac{2}{(m_f^2-q^2_1x(1-x))^
{2-{D\over 2}}}-\frac{1}{(m_f^2-q^2x(1-x))^{2-{D\over 2}}}\right]\right\}~.
\label{z95}
\en
This equation gives the two-loop correction to the gluon trajectory in a
closed form. The summation is performed over quark flavours and $t=q^2$.
Remind that all vectors here are ($D-2$)-dimensional and space-like,
$r^2=-\bar r^2$.

\vskip 0.3cm

{\bf 6. Summary}

We have calculated the two-loop correction $\omega^{(2)}(t)$ to the trajectory
of the Reggeized gluon in QCD. To find this correction we used
the scattering process of massive quarks at large energies $\sqrt{s}$ and fixed
momentum transfer $\sqrt{-t}$. The correction is given by Eq.(\ref{z12}) in
terms of the helicity conserving part of s-channel
discontinuity $\Delta_S$
of the amplitude with colour octet state and negative
signature in the $t$ channel. Furthermore, it depends on the leading
contribution $\omega^{(1)}(t)$ to the trajectory \cite{FKL} (see
Eq.(\ref{z4}))
and the one-loop correction to the helicity conserving part of the quark-
quark-Reggeon vertex $\Gamma^{(+)}_{QQ}(t)$ \cite{FFQ1} (see Eqs.
(\ref{z22})-(\ref{z24}), (\ref{z27}), (\ref{z28}) and (\ref{z31})).
\vskip 0.3cm
The discontinuity $\Delta_S$, obtained in the two-loop approximation,
consists of two contributions, $\Delta^{(2)}_S$ (see Eq.(\ref{z21})) and
$\Delta^{(3)}_S$ (see Eqs.(\ref{z86}), (\ref{z82}),(\ref{z84}) and
(\ref{z73})), which come from two and three particle intermediate states in
the unitary condition.
\vskip 0.3cm
We have obtained the final result~(\ref{z95}) for the two-loop contribution
$\omega^{(2)}(t)$ to the trajectory as a consequence of a series of remarkable
cancellations. These cancellations exhibit the independence of
$\omega^{(2)}(t)$
on any specific property of the scattered quarks and confirm the gluon
Reggeization beyond the leading logarithmic approximation.
\vskip.3cm
Eq.(\ref{z95}) gives us the correction $\omega^{(2)}(t)$ for the space-time
dimension $D$. Of course, our main interest relies in the physical case $D=4$.
Unfortunately, in this case the correction shows up divergences which are
both ultraviolet and infrared. Indeed, the former ones are not difficult to
deal with. They are due to simple first order poles and can be removed by
expressing the bare gauge coupling constant $g$ in terms of the renormalized
one in the total expression for the trajectory:
\eqn
\omega(t) = \omega^{(1)}(t)+\omega^{(2)}(t)+\cdot\cdot\cdot\cdot~.
\enn
In the $\overline{MS}$ scheme one has
\eq
g = g_{\mu}{\mu}^{2-\frac{D}{2}}\left\{1+\left(\frac{11}{3}N-\frac{2}{3}n_f
\right)\frac{g^2_{\mu}}{{(4\pi)}^2}\left[\frac{1}{D-4}-\frac{1}{2}
\ln({4\pi})-\frac{1}{2}\psi(1)\right]+\cdot\cdot\cdot\right\}~,
\label{z96}
\en
where $g_{\mu}$ is the renormalized gauge coupling constant at the
normalization point $\mu$.
\vskip.3cm
The infrared divergences, on the contrary, are much more severe. They could
not be cancelled inside the trajectory $\omega(t)$ because the gluon is a
colour object, whereas we may expect their cancellation for the scattering
of  colourless objects only. So, we should check the cancellation of the
infrared divergences when the trajectory would be put into the Bethe-Salpeter
type equation for the amplitude with vacuum quantum numbers in the $t$
channel. We hope to do that in subsequent papers.

\vskip 1.5cm
\underline {Acknowledgement}: One of us (V.F.) thanks the Dipartimento di
Fisica della Universit\`a della Calabria and the Istituto Nazionale di
Fisica Nucleare - Gruppo collegato di Cosenza for their warm hospitality
while part of this work was done. He thanks also the International Science
Foundation for financial support.

\newpage

\centerline{\bf Figure Captions}
\vskip .3 cm
\begin{description}

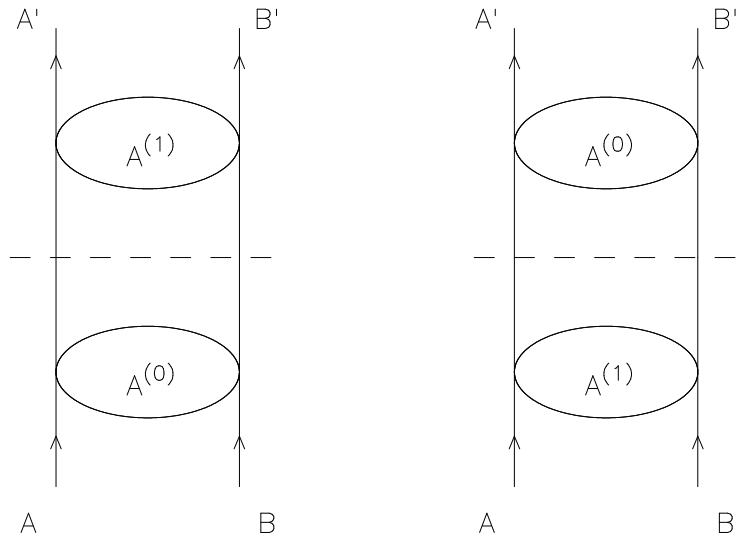
\item{Fig. 1:}
Two particle contribution to the two-loop s-channel
discontinuity. One of the amplitudes is in the Born approximation,
the other one in the
one-loop approximation.
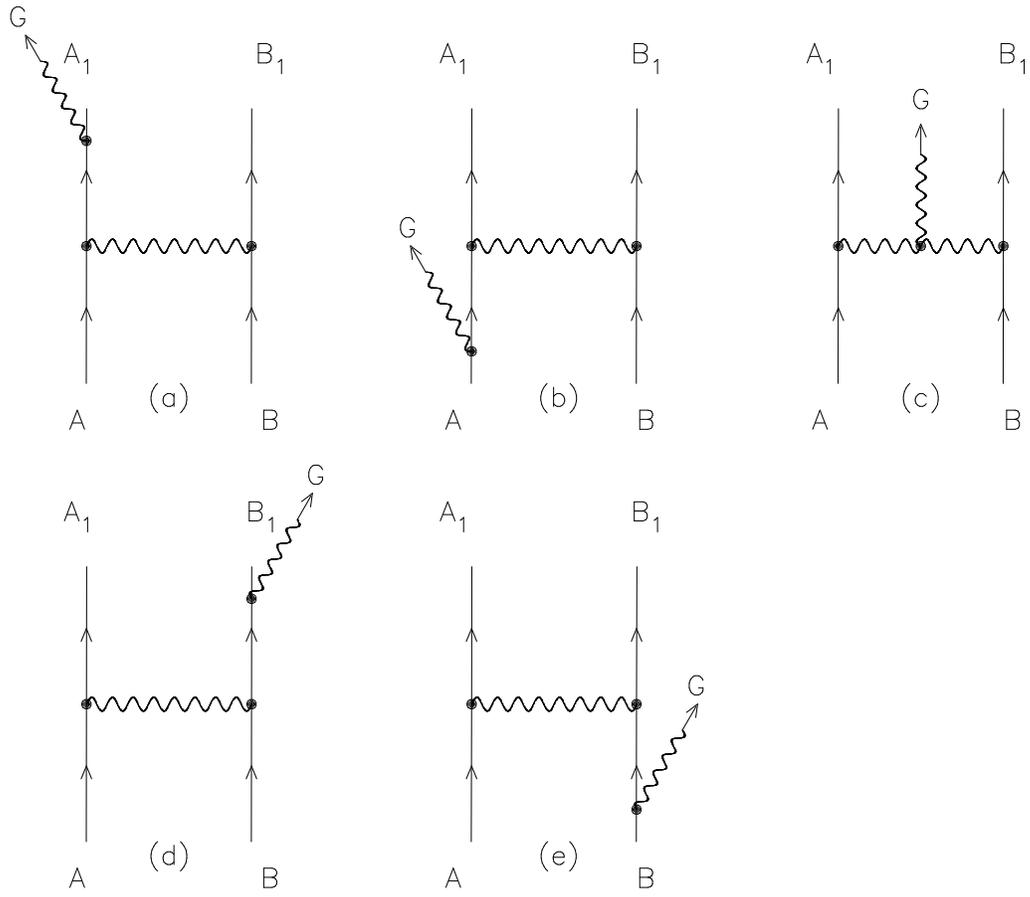
\item{Fig. 2:}
Feynman diagrams for gluon emission.
\end{description}


\begin{thebibliography}{99}

\bibitem{GLR}
L.V. Gribov, E.M. Levin and M.G.Ryskin, Phys. Rep.  {\bf C100} (1983) 1.

\bibitem{AL}
G. Altarelli, Phys. Rep.  {\bf 81} (1982) 1.

\bibitem{FKL}
V.S. Fadin, E.A. Kuraev and L.N.Lipatov, Phys. Lett. {\bf B60} (1975) 50;
E.A.Kuraev, L.N. Lipatov and V.S. Fadin, Zh. Eksp. Teor. Fiz. {\bf 71} (1976)
840 [Sov. Phys. JETP {\bf44} (1976) 443]; {\bf 72} (1977) 377 [{\bf45} (1977)
199].

\bibitem{MLBCN}
A.H. Mueller and J. Qiu, Nucl. Phys. {\bf B268} (1986) 427;
L.N.Lipatov, in "Perturbative Quantum Chronodynamics", ed. A.H. Mueller,
World Scientific, Singapore, 1989;
A.H. Mueller, Nucl. Phys. {\bf B335} (1990) 115;
E.M. Levin, M.G.Ryskin and A.G. Shuvaev, Nucl. Phys. {\bf B387} (1992) 589;
J. Bartels, Phys. Lett. {\bf B298} (1993) 204; Z. Phys. {\bf C60} (1993)
471;
S. Catani and F. Hautmann, Phys. Lett. {\bf B315} (1993) 157;
N.N. Nikolaev, B.G. Zakharov and V.R.Zoller, KFA-IKP(th)-1994-1;
A.H. Mueller, CU-TP-640, 1994.

\bibitem{MNK}
A.H. Mueller and H. Navalet, Nucl. Phys. {\bf B282} (1987) 727;
J. Kwiecinski, A.D. Martin and P.J. Sutton, Phys. Lett. {\bf B278} (1992) 254;
W.K. Tang, Phys. Lett. {\bf B278} (1992) 363;
J. Bartels, A. De Roceck and M. Lowe, Z. Phys. {\bf C54} (1992) 635;
A.D. Martin, W.J. Stirling and R.G. Roberts, Phys. Lett. {\bf B306} (1992)
145.

\bibitem{LF}
L.N. Lipatov and V.S. Fadin, Zh. Eksp. Teor. Fiz. Pis'ma {\bf 49} (1989) 311;
[JETP Lett. {\bf 49},352 (1989);L.N. Lipatov and V.S. Fadin, Yad. Fiz.
{\bf 50} (1989) 1141 [Sov. J. Nucl. Phys. {\bf 50} (1989) 712].

\bibitem{BLF}
Y.Y. Balitskii, L.N. Lipatov and V.S. Fadin, in Materials from the
Fourteenth Winter School of the Leningrad Nuclear Physics Institute [in
Russian], 1979, L; p.109.

\bibitem{FL}
V.S. Fadin and L.N. Lipatov, in {\sl Deep Inelastic Scattering},
Proceedings of the Zeuthen Workshop on Elementary Particle Theory,
Teupitz/Brandenburg, Germany, 1992, edited by J.B. Bl\"umlein and T. Riemann
[Nucl. Phys. {\bf B} (Proc. Suppl.) {\bf 29A} (1992) 93];
V.S. Fadin and L.N. Lipatov, Nucl. Phys. {\bf B406} (1993) 259.

\bibitem{FFQ2}
V.S. Fadin, R. Fiore and A. Quartarolo, Phys. Rev. {\bf D} (1994) 5893

\bibitem{VF}
V.S. Fadin, Budker Institute of Nuclear Physics preprint, BUDKERINP/94-103,
1994.

\bibitem{FF}
V.S. Fadin and R. Fiore, Phys. Lett. {\bf B294} (1992) 286.

\bibitem{FFQ1}
V.S. Fadin, R. Fiore and A. Quartarolo, Phys. Rev. {\bf D} (1994) 2265

\end{thebibliography}
\end{document}